\documentclass[aps,pra,groupedaddress,notitlepage]{revtex4-1}

\usepackage{amssymb}
\usepackage{amsmath}
\usepackage{amsfonts}
\usepackage{amsthm}
\usepackage{graphicx}
\usepackage{color}
\newcommand{\comment}[1]{}

\graphicspath{{figures/}{fig/}{./}}

\begin{document}
\title{Quantifying Non-Stationarity with Information Theory}
\author{Carlos Granero-Belinch\'on $^{1,2}$, St\'ephane G. Roux $^1$  and  Nicolas B. Garnier $^{1,*}$}
\address{%
$^{1}$ Univ Lyon, Ens de Lyon, Univ Claude Bernard, CNRS UMR 5672, Laboratoire de Physique, F-69342 Lyon, France; \\
$^{2}$ IMT Atlantique, Lab-STICC, UMR CNRS 6285, F-29238 Brest, France.}
% Contact information of the corresponding author
%\corres{Correspondence: nicolas.garnier@ens-lyon.fr}

\begin{abstract}
We introduce an index based on information theory to quantify the stationarity of a stochastic process.
The index compares on the one hand the information contained in the increment at the time scale $\tau$ of the process at time $t$ with, on the other hand, the extra information in the variable at time $t$ that is not present at time $t-\tau$. By varying the scale $\tau$, the index can explore a full range of scales. We thus obtain a multi-scale quantity that is not restricted to the first two moments of the density distribution, nor to the covariance, but that probes the complete dependences in the process. 
This index indeed provides a measure of the regularity of the process at a given scale.
Not only is this index  able to indicate whether a realization of the process is stationary, but its evolution across scales also indicates how rough and non-stationary it is.
We show how the index behaves for various synthetic processes proposed to model fluid turbulence, as well as on experimental fluid turbulence measurements.
\end{abstract}

\maketitle 
%\tableofcontents

%%%%%%%%%%%%%%%%%%%%%%%%%%%%%%%%%%%%%%%%%%%%%%%
\section{Introduction}

Many if not most real-world phenomena are intrinsically non-stationary, while most if not all classical tools such as Fourier analysis and power spectrum, correlation function, wavelet transforms, etc., are only defined for---and hence supposed to operate on---signals which are stationary. 
The assumption that a signal or a stochastic process is stationary can either be strict, as in the most formal approaches, or made weaker, as a pragmatic adaptation to the tools used during analysis. The strict stationarity assumption requires all statistical properties, including the probability density function and the complete dependence structure, to be time-invariant. The weak-sense stationarity assumption most commonly used in practice requires the first two moments of the probability distribution to exist and to be time-invariant, as well as the auto-covariance function that is required to be time-translation invariant, which leads to the definition of the correlation function. 

The weak stationarity hypothesis is commonly used to analyze data obtained in various physical, natural, medical or complex systems, in order to apply classical techniques involving the correlation function. While sometimes very well adapted to the data, it may in other situations be a little far-stretched. Let us consider two typical situations which arise, for example, in weather and climate data series: trends and periodic evolutions, {which are known for leading towards long-range dependences~\cite{Beran:2013}, and hence possible non-stationarity}.
For non-stationary signals which present a drift or a trend, a very common and elegant technique consists of time-deriving the signal, and hoping or hypothesizing that the resulting quantity is stationary. If the original trend is not linear in time, a residual trend may still be present in the time-derivative; one can then imagine time-deriving again, iteratively, until the required stationarity assumption is satisfied. Unfortunately, this modus operandi has a drawback, in that it amplifies noise at larger frequencies or smaller scales where it strongly perturbs the power spectrum. As a consequence, it may be difficult to confirm a posteriori whether the iterative time-derivation really gives a stationary process.
For signals that present periodic components, one can restrict the analysis to short time-intervals (examining the weather changes, e.g., temperature fluctuations, over the course of a week should not be impaired by seasonal variations), or on the contrary to long time-intervals (averaging temperature over the course of a year, or heavily sub-sampling  in order to remove any seasonal variation~\cite{Benoit2018}). Unfortunately, this may be extremely reductive and may result in dropping a lot of interesting information located at small scales.

It therefore seems interesting to suggest that the notion of stationarity may depend on the scale at which one is considering the process. Whether one is dealing with epidemiology~\cite{Cazelles2018}, climate~\cite{Cheng2014}, meteorology~\cite{Benoit2018} or animal populations~\cite{Szuwalski2016} among an immense number of possible fields, one might be interested in quantifying the non-stationarity of a dataset depending on the observation scale.

% state of the art ---> testing and index

Identifying and characterizing non-stationarity has always been of utmost importance~\cite{Grenander1957,Priestley1969}. 
Since then, many rigorous techniques have been developed to analyze specific long-range dependences' properties, as can be seen, for example, in~\cite{Beran:2013} for a recent review. To more specifically gauge and quantify non-stationarity, various
approaches have been proposed~\cite{Sachs2000,Dwivedi2011,Dette2011,Barlett2015,Jentsch2015,Cardinally2018} that are based on testing the hypothesis that the process (or sometimes its time-derivative) is stationary with an either positive or negative answer. Depending on the very stationarity hypothesis that is tested, various kinds of non-stationarity are then considered. 
Other approaches have suggested using the roughness of the process, computed in sliding windows, to quantify the order of its non-stationarity~\cite{Das2016}. We proposed following such an approach, but generalizing it on the full range of scales, without restricting it to an appropriate time window.
{The roughness or regularity of a signal is described by its Hurst exponent ${\cal H}$, which can be defined when the power spectrum density of the signal behaves as a power-law of the frequency with an exponent $\alpha$ by asserting $\alpha=-(2{\cal H}+1)$. For example, according to the Kolmogorov K41 theory~\cite{Kolmogorov1991}, the power spectrum of the Eulerian velocity---the kinetic energy spectrum~---in an isotropic and homogeneous turbulent flow behaves as a power law with the exponent $-5/3$, which corresponds to a Hurst exponent $1/3$~\cite{Frisch:1995}. As we discuss in this article, such a power law power spectrum cannot exist in the full range of frequencies for a physical process and it is usually expected that at smaller frequencies---or larger time scales---the process should be stationary.}
In that respect, one could use any method to assess the {roughness} of a signal {and estimate the Hurst exponent~\cite{Gneiting:2012}}, e.g., using the multifractal formalism~\cite{Muzy:1993, Wendt:2009}.

% Our proposal and its benefits (multiscale stationarity, no hypothesis testing, no hypothesis on linearity nor distribution)

In this article, we introduce an index based on information theory to quantify the stationarity of a signal.
Not only is  this index able to indicate whether a realization of the process is stationary at a given scale---typically the size of the realization---but its evolution across scales also indicates how rough and non-stationary the process is.
This index can be interpreted as measuring the extra information contained in the increment of size $\tau$ at time $t$ of the process that is not measured when instead considering the information in the variable at time $t$ that is not present in the variable at time $t-\tau$. By varying the scale $\tau$, the index can explore a full range of scales. As a consequence, the index is a multi-scale quantity. Moreover, it is not restricted to the first two moments of the density distribution, nor to the covariance, but probes the complete dependences in the process. 
We show how the index behaves for various synthetic and real-world processes using fluid turbulence and its diverse landscapes with various scale-invariance properties as the main illustrative theme across our numerical explorations.

% Structure of the article

This article is organized as follows.
In Section~\ref{sec:theory}, we introduce the new stationarity index using information theory. Within the general time-dependent framework and within an appropriately time-averaged framework, we introduce all the building blocks that we then assemble to construct a non-stationarity index. In the limit case of processes with Gaussian statistics and adequate stationarity, we derived analytical expressions for this index. 
In Section~\ref{sec:fGnfBm}, we present our findings on fractional Gaussian noise (fGn), and successive time-integrations of the fGn, which are increasingly non-stationary. We use these Gaussian scale-invariant processes with long-range dependence structures as a set of benchmarks where numerical estimations can be compared with analytical results.
In Section~\ref{sec:physProc}, we focused on synthetic processes that were previously designed to satisfy important physical properties, namely to be stationary at larger scales, as well as smooth enough at smaller scales. We explore how our index 
can characterize non-stationarity depending on the scale on such realistic or physical processes.
In Section~\ref{sec:turbulence}, we use our index to analyze experimental datasets acquired in various fluid turbulence setups, and discuss how such complex real-world data may differ from the synthetic signals of former sections.
Finally, Section~\ref{sec:conclusion} sums up our work and suggests future perspectives.

%%%%%%%%%%%%%%%%%%%%%%%%%%%%%%%%%%%%%%%%%%%%%%%
\section{A Measure of Stationarity and Regularity Using Information Theory}
\label{sec:theory}

This section introduces a novel measure based on information theory to probe the stationarity or the regularity of a discrete-time signal $X$, viewed as a discrete-time stochastic process $X=\left\lbrace x_t \right\rbrace_{t\in\mathbb{N}}$.
After setting up our notations, we recall definitions of time-dependent entropies in the general framework where statistics of the process are considered at a fixed time $t$. We then present the more convenient and practical ``time-averaged framework''~\cite{GraneroBelinchon2019} which is better suited for real-world signals where the number of realizations may be very small. Within this practical framework, entropies are defined using averages over a time window which represents, for example, the time duration of an experiment.
The new stationarity/regularity measure is then defined in both frameworks.

For a discrete-time stochastic process $X=\left\lbrace x_t \right\rbrace_{t\in\mathbb{N}}$, we note $p_{x_t}$ as its probability density function (PDF) at any fixed time $t$, {i.e.}, the PDF of the random variable $x_t$. 
To access the temporal dynamics, we use the Takens time-embedding procedure~\cite{Takens1981} and consider at a given time $t$ the $m$-dimensional vector: 
\begin{equation}
\textbf{x}_t^{(m,\tau)}=\left(x_t, \,x_{t-\tau}, \, \cdots, \, x_{t-(m-1)\tau}\right)\,, \label{eq:def:embed}
\end{equation}
where the time delay $\tau$ is the time scale that we are probing and the embedding dimension $m$ controls the order of the statistics which are explicitly involved. 
We note $X^{(m,\tau)}=\{\textbf{x}_t^{(m,\tau)}\}_{t\in\mathbb{N}}$ as the corresponding stochastic process at the time scale $\tau$.

In addition to the time-embedding procedure, we also consider increments of the signal $X$ at time-scale $\tau$. At a given time $t$, such an increment reads:
\begin{equation}\delta_{\tau}x_t=x_t-x_{t-\tau} \,, \label{eq:def:inc}
\end{equation}
and we define the stochastic process $\delta_\tau X=\left\lbrace \delta_\tau x_t \right\rbrace_{t\in\mathbb{N}}$ at the time scale $\tau$.

We use in this article the differential entropy for continuous processes, although all results presented here hold for discrete processes, by using the Shannon entropy. Given a probability density function (PDF) $p$, the entropy is a functional of $p$:
\begin{equation}
H = -\int_{x\in{\mathbb R}} p(x) \ln(p(x)) {\rm d}x \,. \label{eq:def:entropy}
\end{equation}

Given a process $X$, we define below various entropies or combinations of entropies of various PDF of random variables pertaining to either increments (\ref{eq:def:inc}) or time-embedded vectors (\ref{eq:def:embed}).
The information theory quantities that we discuss below for $X$ thus depend on the time-scale $\tau$; varying the time-scale $\tau$ allows a multi-scale analysis of the process dependences.

%%%%%%%%%%%%%%%%%%%%%%%%%%%%%%%%%%%%%%%%%%%%%%%%%%%%%%%%%%%%%%%%%
\subsection{General Framework}

We recall here how one can define entropies for any stochastic process $X$, whether $X$ is stationary or non-stationary. Because the PDF of the random variable $x_t$ {\em a priori} depends on time $t$, each random variable is considered separately.
Within this very general framework, different entropies are defined for the process $X$ at each time step $t$.

\subsubsection{{Shannon} Entropy of the Time-Embedded Process}
%we change this as subsubsection, please confirm, and check all similar places

We define $H_t(X^{(m,\tau)})$, the entropy of the time-embedded process $X^{(m,\tau)}$ at time $t$, using the entropy formula (\ref{eq:def:entropy}) for the $m$-dimensional multivariate PDF $p_{\textbf{x}_t^{(m,\tau)}}$ of the random variable $\textbf{x}_t^{(m,\tau)}$:

\begin{equation}
H_t(X^{(m,\tau)})\equiv H(\textbf{x}_t^{(m,\tau)}) \equiv-\int_{\mathbb{R}^{m}} p_{\textbf{x}_t^{(m,\tau)}}(\textbf{x}) \ln(p_{\textbf{x}_t^{(m,\tau)}}(\textbf{x})) {\rm d}\textbf{x}
\label{eq:entropy:abstract}
\end{equation}

This quantity depends on the time $t$ at which the process is considered, as well as on the time scale $\tau$ involved in the embedding procedure. 
We simply note it $H_t^{(m,\tau)}(X)$ for the signal $X$ under consideration. 

The entropy $H_t^{(m,\tau)}(X)$ involves the complete PDF of the variable $\textbf{x}_t^{(m,\tau)}$, including high-order moments. Therefore, it depends on high-order statistics. Nevertheless, it does not depend on the first-order moment and any random variable can be centered without altering its entropy. 

For $m=1$ (no embedding), the entropy does not depend on $\tau$ nor on the dynamics of the process $X$; in that specific case, we simply note it $H_t(X)$. 
As soon as $m>1$, the entropy $H_t^{(m,\tau)}(X)$ depends on the complete dependence structure of the components of the embedded vector $\textbf{x}_t^{(m,\tau)}$, and hence, $H_t^{(m,\tau)}(X)$ probes the linear and non-linear dynamics of the process at scale $\tau$ and time $t$.

\subsubsection{Shannon Entropy of the Increments}

We define $H_t(\delta_\tau X) \equiv H(\delta_\tau x_t)$ as the entropy of the increments process $\delta_\tau X$ at time $t$ by applying the definition (\ref{eq:def:entropy}) to the PDF of the random variable $\delta_\tau x_t$.

\subsubsection{Entropy Rate}

We define $h_{t}^{(m,\tau)}(X)$, the entropy rate of order $m$ at time $t$ and at time-scale $\tau$ of the process $X$, as the variation of Shannon entropy between the random variables $\textbf{x}^{(m,\tau)}_{t-\tau}$ and $\textbf{x}^{(m+1,\tau)}_{t}$, {i.e.}, the increase in information from two successive time-embedded versions of the process $X$ at time $t$:
\vspace{-12pt}

\begin{subequations}
\begin{align}
h_{t}^{(m,\tau)}(X) &\equiv H(\textbf{x}_t^{(m+1,\tau)}) - H(\textbf{x}_{t-\tau}^{(m,\tau)}) \nonumber \\
&= H_t^{(m+1,\tau)}(X) - H_{t-\tau}^{(m,\tau)}(X) \label{eq:h:abstract:diff} \\
&=  H_t(X) - I_{t}^{(m,\tau)}(X) \,, \label{eq:h:abstract:MI}
\end{align}
\end{subequations}
%please do not use sub-equations in the main text
%
where the auto-mutual information $I_{t}^{(m,\tau)}$ is the mutual information between the two random variables $x_t$ and $\textbf{x}_{t-\tau}^{(m, \tau)}$ which together form the $m+1$ time-embedded variable $\textbf{x}_t^{(m+1, \tau)}$:
%\begin{subequations}
\begin{align}
I_{t}^{(m,\tau)}(X) &\equiv H(x_{t}) + H(\textbf{x}^{(m,\tau)}_{t-\tau}) - H(x_t, \textbf{x}_{t-\tau}^{(m,\tau)}) \nonumber \\
&\equiv H_{t}(X) + H^{(m,\tau)}_{t-\tau}(X) - H^{(m+1,\tau)}_t(X) \,. \label{eq:MI} 
\end{align}
%\end{subequations}

For non-stationary processes, $I^{(m,\tau)}_t$ offers a generalization of the auto-covariance. For stationary processes, $I^{(m,\tau)}_t$ is independent on the time $t$ and is a generalization of the auto-correlation function~\cite{GraneroBelinchon2019a}.

In the remainder of this article, we focus on the entropy rate of order $m=1$, which we note $h_t^{(\tau)}$.

%%%%%%%%%%%%%%%%%%%%%%%%%%%%%%%%%%%%%%%%%%%%%%%%%%%%%%%%%%%%%%%%%
\subsection{Time-Averaged Framework}\label{sec:taframework}

When a single realization of a process $X$ is available, we assume some form or ergodicity and treat the set of values $x_t$ as realizations of a stationary process. 
This crude assumption is indeed fruitful, and very convenient when a single signal or a single time series is available.
Let us note by $[t_0, t_0+T[$ the time window of length $T$ corresponding to the available realization of $X$. We consider the probability density function $\bar p_{T,t_0,x}$ obtained by considering all data points within the time window~\cite{GraneroBelinchon2019}. Since this quantity is a time-average, it does not explicitly depend on time $t$ but on the total duration $T$ and on the starting time $t_0$. 

Considering the time-embedded process $X^{(m,\tau)}=\{\textbf{x}_t^{(m,\tau)}\}_{t\in\mathbb{R}}$, the time-averaged PDF can be expressed as
\begin{equation}
\bar p_{T, t_0, \textbf{x}^{(m,\tau)}}(\textbf{x}) = \frac{1}{T} \sum_{t=t_{0}}^{t_{0}+T-1} p_{\textbf{x}^{(m,\tau)}_{t}}(\textbf{x}) %{\rm d}t 
\end{equation}

For a stationary process, $\bar{p}_{T, t_0, \textbf{x}^{(m,\tau)}} = p_{\textbf{x}_t^{(m,\tau)}}$: the time-averaged PDF does not depend on $T$ or $t_0$ and matches the stationary PDF of the process $X$.
Using time-averaged PDFs for any process, we define {\em ersatz} versions of the entropies presented in the previous section as follows.

\subsubsection{{Shannon} Entropy}
We define the ersatz entropy $\bar{H}^{(m,\tau)}_{T}(X)$ of the signal $X$ in the time window $[t_0, t_0+T[$ as the entropy (\ref{eq:def:entropy}) of the time-averaged PDF $\bar p_{T,t_0, \textbf{x}^{(m,\tau)}}$ of the time-embedded process $X^{(m,\tau)}$:

\begin{equation}\label{eq:HT}
\bar{H}_{T}^{(m,\tau)}(X)=-\int_{\mathbb{R}^{m}} \bar p_{T,t_0, \textbf{x}^{(m,\tau)}}(\textbf{x}) \ln(\bar p_{T,t_0, \textbf{x}^{(m,\tau)}}(\textbf{x})) {\rm d}\textbf{x}
\end{equation}

This entropy $\bar{H}_{T}^{(m,\tau)}(X)$ describes the complexity of the set of all successive values of the process $X^{(m,\tau)}$ in the time interval $[t_0, t_0+T[$. It can be interpreted as the amount of information needed to characterize the available realization of the process in the time interval $[t_0, t_0+T[$. It depends on $T$ and $t_0$ but in order to simplify the notations, we drop the index $t_0$ in the following.

\subsubsection{{Entropy} of the Increments}
We define $\bar{H}_T(\delta_\tau X)$, the ersatz entropy of the increments of the signal $X$ at the time scale $\tau$ in the time window $[t_0, t_0+T[$, as the entropy (\ref{eq:def:entropy}) of the time-averaged PDF of the increment process $\delta_\tau X$.

\subsubsection{{Entropy} Rate}

We define the ersatz entropy rate $\bar h_{T}^{(m,\tau)}(X)$ of the signal $X$ in the time window $[t_0,t_0+T[$ as the increase in ersatz entropy when increasing the embedding dimension by +1. This is thus the same expression as in the general framework but using time-averaged probabilities along the trajectory of the process:
\begin{subequations}
\begin{align}
\bar h_{T}^{(m,\tau)}(X) 
&= \bar{H}_{T}^{(m+1,\tau)}(X) - \bar{H}_{T}^{(m,\tau)}(X) \,, \label{eq:hT} \\
&=  \bar H_{T}(X) - \bar I_{T}^{(m,1,\tau)}(X) \label{eq:hT2}
\end{align}
\end{subequations}

%%%
For non-stationary processes with centered stationary increments, $t_0$ only influences the mean of the distribution; all centered moments only depend on $T$, the size of the time-window. Therefore, in this case, all information quantities only  depend on $T$.

%%%%%%%%%%%%%%%%%%%%%%%%%%%%%%%%%%%%%%%%%%%%%%%%%%%%%%%%%%%%%%%%%
\subsection{Towards a Measure of Regularity and Stationarity}

Exploring the dynamics along scales $\tau$ of a signal, viewed as a stochastic process, can be achieved with information theory in two distinct ways with the tools presented above. The first one is to consider the increments and compute their entropy. The second one is to consider the time-embedding and hence use the entropy rate. Both naturally introduce the time-scale $\tau$ and are able to probe the dependences between two variables of the process separated by $\tau$.

On the one hand, the entropy of the increments measures the uncertainty---or information---in the increment which represents the variation between $x_{t-\tau}$ and $x_{t}$. This approach is appropriate for signals which are not stationary but have stationary increments. It thus also offers a direct comparison with traditional tools which heavily rely on the use of increments to analyze signals. For example, Ref. \cite{GraneroBelinchon2018} used the entropy of the increments to examine a variety of synthetic multi-fractal processes together with experimental velocity measurements in fully developed turbulence.

On the other hand, the entropy rate ($h_t^{(m,\tau)}$ or $\bar{h}_T^{(m,\tau)}$) measures the amount of uncertainty---or new information---in the extra variable $x_{t}$ that is not already accounted for when considering the variable $x_{t-\tau}$. As such, it can be viewed as a measure of the dependences at scale $\tau$. For example, in the case of stationary signals, the entropy rate can be used to characterize the scale-invariance of fully developed turbulence~\cite{GBelinchon2016} or to probe higher order dependences beyond mere second-order correlations~\cite{GraneroBelinchon2019a}.

Both the entropy of the increments and the entropy rate can be computed in the time-averaged framework presented in Section~\ref{sec:taframework}.
Interestingly, for non-stationary processes with stationary increments, both measures are almost stationary, {i.e.}, they only weakly depend  on  the time-interval length~$T$~\cite{GraneroBelinchon2019}. While this property is expected for the entropy of the increments which are stationary---so $H_{T}(\delta_\tau X)=H_t(\delta_\tau X)$ does not depend on $T$ or $t$---this is more surprising for the entropy rate.
This illustrates that the entropy of the increments and the entropy rate are not identical at all, albeit both exploring the dynamics between $x_t-\tau$ and $x_t$. With this in mind, we propose using the difference between these two information quantities as an index to finely probe the non-stationarity of a process.

\subsubsection{Relation between $h_t^{(\tau)}(X)$ and $H_t(\delta_{\tau}X)$ in the General Framework}

Given a non-stationary process $X$, we define the index:
\begin{align}
\Delta_t^{\tau}(X) &\equiv H_t(\delta_\tau X) - h_t^{(\tau)}(X) \,. \label{eq:Delta:abstract}
\end{align}

We can rewrite $\Delta_t^{\tau}$ by first expressing the entropy of $X^{(2,\tau)}$ at time $t$:
\begin{align}
H^{(2,\tau)}_t(X) \equiv H(\textbf{x}_t^{(2,\tau)}) &=H\left( x_t, x_{t-\tau}\right) \nonumber \\
&= H \left(x_t, \delta_\tau x_t \right) \,. \label{eq:H2:inc}
\end{align}

This follows from writing $x_t$ as the sum $\delta_\tau x_t + x_{t-\tau}$ and using chained conditioned probabilities.
According to Equation (\ref{eq:h:abstract:diff}), the entropy rate of order 1 then reads:
\begin{align}
h^{(\tau)}_t(X) &\equiv H^{(2,\tau)}_t(X) - H^{(1,\tau)}_t(X) \nonumber \\
&= H(x_t, \delta_\tau x_t) - H(x_t) \nonumber \\
%&= H(\delta_\tau x_t | x_t) \\
&= H(x_t, \delta_\tau x_t) - H(x_t) - H(\delta_\tau x_t) + H( \delta_\tau x_t) \nonumber \\
&= - MI( x_{t-\tau}, \delta_\tau x_t) + H_t(\delta_{\tau} X)   \label{eq:h_vs_Hinc}\,,
\end{align}
where $MI(X,Y)\equiv H(X)+H(Y)-H(X,Y)$ is the mutual information between the signals $X$ and $Y$, here the variable $x_{t-\tau}$ and the increment $\delta_\tau x_t$ leading from $x_{t-\tau}$ to $x_t$.
This relation holds for any process; in particular, the stationarity of the increments is not required. This leads to:
\begin{align}
\Delta_t^{\tau}(X) &= MI( x_{t-\tau}, \delta_\tau x_t) \ge 0 \,. \label{eq:Delta:abstract:MI} 
\end{align}
where $\Delta_t^{\tau}$ is a combination of three entropies that can be rewritten as a mutual information; therefore, it is always greater than or equal to 0.

By definition, (\ref{eq:Delta:abstract}) $\Delta_t^{\tau}$ quantifies the extra information---or extra uncertainty---which is present in the increment $\delta_\tau x_t = x_t - x_{t-\tau}$ but is not accounted for when measuring the increase in information between $x_{t}$ and $(x_t, x_{t-\tau})$. Then, the rewriting into (\ref{eq:Delta:abstract:MI}) shows that $\Delta_t^{\tau}$ also corresponds to the shared information between the walk $X$ at time $t-\tau$ and the next increment $\delta_\tau$ that leads to the walk at time $t$. In other words, $\Delta_t^{\tau}$ is the difference between on the one hand the sum of the information contained in $x_t$ and the information contained in the increment $x_{t}-x_{t-\tau}$, and on the other hand the information in the vector $(x_t,x_{t-\tau})$. 
Both interpretations clearly illustrate that, although the information in the vectors $(x_t, x_{t-\tau})$ and $(x_t, \delta_\tau x_t)$ is the same (see Equation (\ref{eq:H2:inc})), the information in $x_t$ cannot be obtained by combining the information of the process at time $t-\tau$ together with 
the information in the increments between the two times $t-\tau$ and $t$.

\subsubsection{Definition of an Index in the Time-Averaged Framework}

The two terms in the right-hand side of Equation (\ref{eq:Delta:abstract}) have counterparts in the time-averaged framework.
We thus define, for any process $X$ indexed on a time-interval of length $T$:
\begin{subequations}
\begin{align}
\bar{\Delta}_T^{\tau}(X) &\equiv \bar{H}_T(\delta_\tau X) - \bar{h}_T^{(1,\tau)}(X) \label{eq:Delta:diff} \\
&= \overline{MI}_T( x_{t-\tau}, \delta_\tau x_t) \label{eq:Delta:MI}\,.
\end{align}
\end{subequations}

We show in the following how this quantity can be used to probe the non-stationarity of a signal under realistic conditions, {i.e.}, when one can only compute entropies in the time-averaged framework, e.g., when a single realization is available.
We further refer to $\bar{\Delta}_T^{\tau}(X)$ as the stationarity or regularity index.

\subsubsection{Expression for a Stationary Process with Gaussian Statistics}
\label{sec:stationary:Gaussian}

All information quantities considered here do not depend on the first moment of the process, which we now consider the zero-mean without loss of generality.
For a process with Gaussian statistics, the dependence structure can be expressed using only the covariance.
As a consequence, all terms in Equation (\ref{eq:Delta:diff}) can be written in terms of the covariance.

Further assuming a stationary process $X$, and noting $\sigma_x$ and $c(\tau)$, its time independent standard deviation and correlation function, we have:
\begin{align}
\bar{H}_T(\delta_\tau X) &= H_t(\delta_\tau X)=\frac{1}{2}\ln(2 \pi e \sigma_{\delta_{\tau}}^2) \,, \label{eq:G:H_inc} \\
\bar{h}_T^{(\tau)}(X) &= h_t^{(\tau)}(X) =\frac{1}{2}\ln(2 \pi e \sigma_{x}^2)+\frac{1}{2}\ln \left(|\Sigma|\right) \,, \label{eq:G:h}
\end{align}
where $\Sigma$ is the correlation matrix of the process $X$ and $\sigma_{\delta_\tau}^2=2\sigma_x^2(1-c(\tau))$ is the variance of its increments $\delta_\tau X$ at scale $\tau$. Using $|\Sigma|=1-c(\tau)^2$ and plugging Equations (\ref{eq:G:H_inc}) and (\ref{eq:G:h}) into Equation (\ref{eq:Delta:diff}) gives:
\begin{equation}
\bar{\Delta}_T^{\tau}(X)={\Delta}_t^{\tau}(X) = \frac{1}{2}\ln\left(\frac{2}{1+c(\tau)}\right) \,. 
\label{eq:G:Delta}
\end{equation}

Thus, the index $\bar{\Delta}_T^{\tau}(X)$ of a stationary process $X$ does not depend on the standard deviation of $X$.

In the specific case of an uncorrelated Gaussian process, the index takes the special value $\bar{\Delta}_T^{\tau}=\ln\sqrt{2}$.
For positive correlations $c(\tau)\ge0$, the index is smaller: $\bar{\Delta}_T^{\tau}(X)\le \ln\sqrt{2}$ while for anti-correlations $c(\tau)\le0$ the index is larger $\bar{\Delta}_T^{\tau}(X)\ge \ln\sqrt{2}$. These results hold for any stationary Gaussian process. 

When the correlation is small, $c(\tau)\ll 1$, Equation (\ref{eq:G:Delta}) can be Taylor-expanded as
\begin{equation}
\bar{\Delta}_T^{\tau}(X)= \ln\sqrt{2} - \frac{c(\tau)}{2} \,. 
\label{eq:G:Delta:exp}
\end{equation}

If we further assume that the process exhibits some self-similarity such that the variance $\sigma_{\delta_\tau}^2$ of its increments  behaves as a power law of the scale $\tau$ with the exponent $\zeta_2$, {i.e.}, $1-c(\tau) \propto \tau^{\zeta_2}$, then taking the logarithm of Equation (\ref{eq:G:Delta:exp}) leads to $\ln {\Delta}_T^{\tau}(X) \propto \zeta_2 \ln \tau$, up to an additive constant.

\subsection{Estimation Procedures for Information Theory Quantities}

All results reported in the present article were computed using nearest neighbors ($k$-nn) algorithms: from Kozachenko and Leonenko~\cite{Kozachenko1987} for the entropy, and from Kraskov, St\"ogbauer and Grassberger~\cite{Grassberger2004} for the mutual information estimator in Equations (\ref{eq:hT2}) and  (\ref{eq:Delta:MI}). These estimators have small bias and small standard deviation~\cite{Grassberger2004, Gao2016, GraneroBelinchon2019, GraneroBelinchon2019a}. Additionally, for each value of the time scale $\tau$, we subsample the available data to eliminate the contribution of dependences from scales smaller than $\tau$~\cite{Theiler1986}.

To have a better comparison between various processes, we always use realizations of the same size $T$, and normalize each realization so that the unit-time increments $(\tau=1)$ have a standard deviation equal to 1. This removes the trivial dependence of the entropy rate on the standard deviation, while it does not affect the index which does not depend on the standard deviation of the process.

\section{fGn and fBm Benchmarks}
\label{sec:fGnfBm}

We focus in this section on fractional Gaussian noise (fGn) and fractional Brownian motion (fBm) which we use as benchmarks for our analysis.
These two processes have Gaussian statistics and are hence easy to analytically manipulate.
They have well-known scale-invariant covariance structures~\cite{Mandelbrot1968} and are commonly used as toy models for systems exhibiting self-similarity and long-range dependences~\cite{Kolmogorov1991}, as observed in, e.g., the vicinity of the critical point in phase transition, or geophysical processes~\cite{Pelletier1999}. 

Historically, the fBm was introduced prior to the fGn: the latter was studied as the derivative of the former~\cite{Mandelbrot1968}. The fBm is widely used in the literature as a prototype walk exhibiting self-similarity and as a natural generalization of the Brownian motion. For clarity, we start our presentation with the fGn which is stationary, and introduce the fBm as a time-integration of the fGn; we also present the process obtained by further time-integrating the fBm.

\subsection{Definitions and Analytical Expressions}

\subsubsection{Fractional Gaussian Noise} 
The fGn $W\equiv\{w_t\}_{t\in\mathbb{N}}$ is a stationary stochastic process with Gaussian statistics and long-range dependences, whose correlation function is expressed as 

\begin{equation}
c_{W}(\tau) = \frac{\sigma^2_{0}}{2} \left[ (\tau-1)^{2\mathcal{H}} -2 \tau^{2\mathcal{H}} + (\tau+1)^{2\mathcal{H}}  \right] \,, \label{def:fBm:correlation}
\end{equation}

\noindent where the prefactor $\sigma_0$ is the standard deviation of the fGn and $1-\mathcal{H}$ is the Hurst exponent~\cite{Mandelbrot1968} (this convention allowing for a direct identification with the fBm defined below). Without loss of generality, we impose $w_0=0$ so that the first value is 0 at time $t=0$.  

Since the fGn is stationary with Gaussian statistics, its non-stationarity index $\bar{\Delta}_T^{\tau}$ is straightforwardly given by Equation (\ref{eq:G:Delta}) with the expression (\ref{def:fBm:correlation}) of the correlation of the fGn.

\subsubsection{Time Integration}

Given a discrete-time stochastic process $X\equiv\{x_t\}_{t\in {\mathbb N}}$ with $x_0=0$, we can define a new stochastic process $Y\equiv\{y_t\}_{t\in {\mathbb N}} = {\mathcal I}(X)$ representing the integration of $X$ over time as
\begin{equation}
%y_0 = 0, \qquad 
y_t = \sum_{t'=0}^{t} x_{t'} \quad \forall t\ge 1\,. \label{eq:builtmotion}
\end{equation}
$Y$ is the {\em motion} or {\em walk} built on $X$. In fact, the process constituted of the increments of $Y$ at scale $\tau=1$ is nothing but $X$.

In all generality, for a continuous-time process, (\ref{eq:builtmotion}) is to be replaced by a continuous integration. Then, $Y$ is a non-stationary process which is more regular than $X$: if $X$ is $n$-differentiable, then $Y$ is $(n+1)$-differentiable. We also note that if $X$ has no oscillating singularity and a Hurst exponent ${\cal H}$, then $Y$ has a Hurst exponent ${\cal H}+1$~\cite{Muzy:1993,Wendt:2009,Abry:2011}. Performing time-integration increases the Hurst exponent by +1 and gives a new process which is ``more non-stationary''.

\subsubsection{Fractional Brownian Motion} 
\label{sec:theory:fBm}

The fBm $B\equiv\{b_t\}_{t\in{\mathbb N}}$ can be defined as the integration over time of the fractional Gaussian noise $W$ as $B={\mathcal I}(W)$. Although fBm with the Hurst exponent ${\cal H}$ is non-stationary, its power spectral density can be defined~\cite{Flandrin1989}; it is a power law of the frequency with exponent $-(2{\cal H}+1)$. The covariance structure of the fBm is given by
\begin{equation}
\mathbb{E} \{b_t b_{t+\tau} \} =  \frac{\sigma^2_0}{2} \left[t^{2\mathcal{H}}+ (t+\tau)^{2\mathcal{H}}-|\tau|^{2\mathcal{H}}\right] \,,
\label{eq:fBm:covariance}
\end{equation}
where $\sigma_0$, the standard deviation of the fBm at unit-time $t=1$ is the standard deviation of the fGn.

In the time-averaged practical framework: we separately consider the two terms\linebreak in (\ref{eq:Delta:diff}). The increments of the fBm are stationary and their standard deviation is $\sigma_0\tau^{\cal H}$. We note $H_0 = \frac{1}{2}\ln \left( 2\pi e \sigma_0^2\right)$ as the entropy of the fGn. The ersatz entropy of the increments of the fBm equals the entropy of the increments in the general framework which is time-independent:
\begin{equation}
\bar{H}_{T}(\delta_\tau B) = H_{t}(\delta_\tau B) = H_0 + \mathcal{H}\ln(\tau) \,. \label{eq:fBm:ersatz:H_inc}
\end{equation}

The ersatz entropy rate cannot be simply expressed but it was shown~\cite{GraneroBelinchon2019} that in the limit $\tau \ll T$:
\begin{equation}
\bar{h}_{T}^{(\tau)}(B) \simeq H_0 + \mathcal{H} \ln \tau - {\cal C}\left( \frac{\tau}{T}\right) \,, \label{eq:fBm:ersatz:h}
\end{equation}
where ${\cal C}\left( \frac{\tau}{T}\right)$ is a correction in $\tau/T$ that depends on ${\cal H}$. Subtracting (\ref{eq:fBm:ersatz:h}) from (\ref{eq:fBm:ersatz:H_inc}), we deduce that the index $\bar{\Delta}_T^{\tau}$ of the fBm vanishes as $\tau/T$ when the duration $T$ of the signal is increased or the time scale $\tau$ is reduced.

\subsubsection{Time-Integrated fBm} We also present below results obtained for $A={\cal I}(B)$, the process obtained by time-integrating the fBm with Equation (\ref{eq:builtmotion}).
Although the covariance structure of this non-stationary process with non-stationary increments is out of the scope of the present paper, we note that its power spectral density is a power law with the exponent $-(2{\cal H}+3)$ while its generalized Hurst exponent is ${\cal H}+1$.

%%%% scale invariant process with Gaussian statistics
\subsection{Numerical Observations}

In this section, we report numerical measurements of ersatz entropies on an fGn, a fBm and a time-integrated fBm obtained with Equation (\ref{eq:builtmotion}). For each of these three processes, 100 realizations with fixed $T=2^{16}$ samples were used. The time scale $\tau$ is varied from $\tau=1$ to $\tau=2^9$. For a given $\tau$, the processes are sub-sampled and one sample is kept for every $\tau$ samples. Consequently, the effective number of points used for the entropies' estimation decreases as $T/\tau$ so the bias and standard deviation are expected to increase with $\tau$ for a fixed $T$~\cite{GraneroBelinchon2019a,GraneroBelinchon2019}.

\begin{figure}[htb]
\includegraphics[width=.9\linewidth]{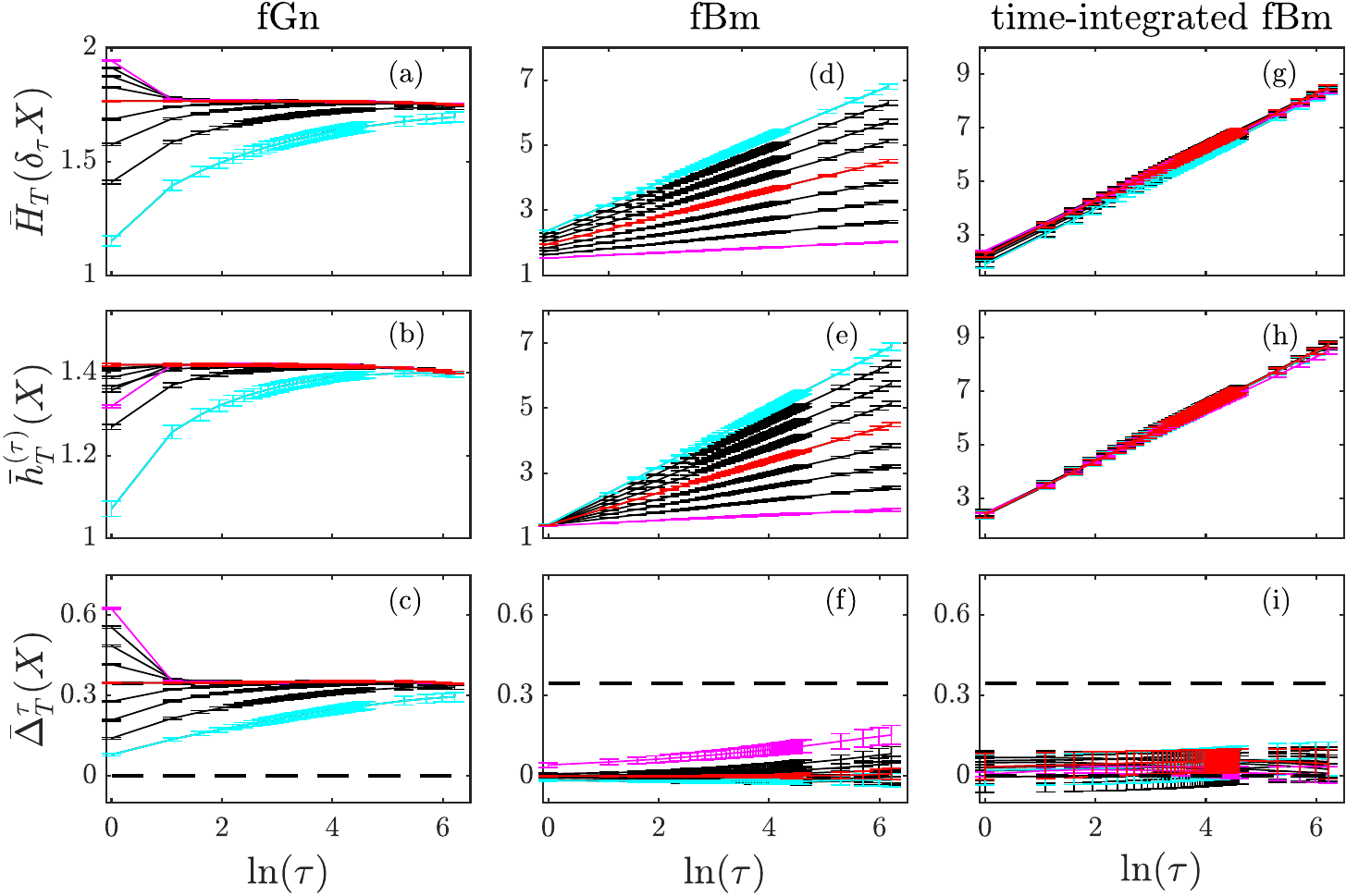}
\caption{{\bf Scale-invariant processes.} %Is the bold necessary? % yes, title is easier to see
Entropy of the increments $\bar{H}_T(\delta_\tau X)$ (\textbf{first line}), entropy rate $\bar{h}^{(\tau)}_T(X)$ (\textbf{second line}) and index $\bar{\Delta}_T^{\tau}(X)=\bar{H}_T(\delta_\tau X)-\bar{h}^{(\tau)}_T(X)$ (\textbf{third line}) for three scale-invariant processes $X$ with various levels of stationarity: fGn (\textbf{first column}), fBm (\textbf{second column}) and time-integrated fBm (\textbf{third column}). For each process, various Hurst exponents ${\cal H}$ in the range $[0.1, 0.9]$ are reported, {with results colored in magenta for ${\cal H}=0.1$, in red for ${\cal H}=0.5$, in cyan for ${\cal H}=0.9$, and in black for other values.
In the third line, special values 0 and $\ln\sqrt{2}$ for the index are represented with horizontal black dashed lines.}
}
\label{fig:FBM}
\end{figure}

Figure~\ref{fig:FBM} presents our results for the three processes: fGn (first row), fBm (second row) and time-integrated fBm (third row) for various Hurst exponents ${\cal H}$ in the range $[0.1, 0.9]$. 
For each process, the entropy of the increments (first line of Figure~\ref{fig:FBM}) and the entropy rate (second line of Figure~\ref{fig:FBM}) exhibit similar behaviors when the time-scale $\tau$ is varied.
For the fGn, these two quantities converge to a constant value when $\tau$ is increased, but it can be seen that the entropy of the increments converges from above when $H<1/2$. For the fBm, the two quantities increase linearly in $\ln\tau$, with a slope that is exactly the Hurst exponent ${\cal H}$~\cite{GBelinchon2016,GraneroBelinchon2018}.
For the time-integrated fBm, which has a generalized Hurst exponent larger than 1, the two quantities also evolve linearly in $\ln\tau$, but with a constant slope 1 independent on ${\cal H}$. This indicates that neither the entropy of the increments nor the entropy rate can be used to estimate ${\cal H}\ge 1$. 

The index ${\Delta}_T^{\tau}$ (third line of Figure~\ref{fig:FBM}) shows a different behavior when $\tau$ is increased. For the fGn (Figure~\ref{fig:FBM}c), it converges to the constant value $\ln\sqrt{2}$ (represented by a horizontal dashed line). This specific value corresponds to the one obtained for stationary Gaussian process that is uncorrelated, which is asymptotically the case for the fGn when $\tau \rightarrow \infty$. We note that ${\Delta}_T^{\tau}$ is exactly $\ln\sqrt{2}$ for the random noise (fGn with $\mathcal{H}=1/2$, uncorrelated, in red in Figure~\ref{fig:FBM}c), while ${\Delta}_T^{\tau}$ converges to $\ln\sqrt{2}$ from above for $\mathcal{H}<1/2$ (negative correlation, curves between magenta and red) and from below for $\mathcal{H}>1/2$ (positive correlation, longer range, curves between red and cyan). All these observations are in perfect agreement with our findings in Section~\ref{sec:stationary:Gaussian}, and in particular with the expression (\ref{eq:G:Delta}).

For the fBm, $\bar{\Delta}_T^{\tau}$ is very close to zero to most values of ${\cal H}$, although a little increase is observed for ${\cal H}< 0.5$. This is in agreement with our findings in Section~\ref{sec:theory:fBm}: the index $\bar{\Delta}_T^{\tau}$ behaves as $\tau/T$ with a prefactor that depends on ${\cal H}$.

For the time-integrated fBm, $\bar{\Delta}_T^{\tau}$ is constant and zero within the error-bars, which are large (Figure~\ref{fig:FBM}i). Larger error-bars are expected on ersatz quantities of processes which are increasingly non-stationary: time-averages along a single trajectory depend more and more on the trajectory. Nevertheless, for such processes, $\bar{\Delta}_T^{\tau} \simeq 0$ which suggests that the quantity of information contained in the increment $\delta_\tau x_t$ is roughly the same as the extra information in $x_t$ with respect to the information in $x_{t-\tau}$.

%%%%%%%%%%%%%%%%%%%%%%%%%%%%%%%%%%%%%%%%%%%%%%%
\section{Physical Stochastic Processes with Dissipative and Integral Scales}
\label{sec:physProc}

The fractional Brownian motion, just as the traditional random walk, is not a physical process encountered ``as is'' in nature, but a mathematical model with at least two drawbacks. Firstly, the power spectrum of the fBm behaves as a power law with an exponent $2{\cal H}+1$, which implies that for ${\cal H}<1/2$, it has an infinite energy in the continuous limit. This is not usually a problem with discrete time, as the sampling frequency is finite. Secondly, {in many non-stationary processes,} the standard deviation diverges with time; {this is for example the case if the process is scale invariant, such as the fBm}. This is again not a problem as any realization under consideration has a finite duration. These two drawbacks are indeed related to the assumption of a perfect scale-invariance of the process in an infinite range of scales; whereas in a physical system, scale invariance is restricted to a finite range of scales only.

Introducing a high frequency cutoff or equivalently a small, or {\em dissipative}, %Is the italics necessary?,% yes for definition
scale $\epsilon$ is a common and natural way to prevent the divergence of the power spectrum; we refer to such an introduction as ``{\em regularization}''~\cite{Pereira2016} in this article. It also offers an interesting perspective to model the behavior of a physical system at smaller scales where the scale invariance property does not hold anymore. Introducing a large, or {\em integral}, scale ${\cal T}$ is a natural way to prevent the divergence of the standard deviation of the process. Interestingly, this also leads to a ``{\em stationarization}'' of the process at scales equal to or larger than ${\cal T}$~\cite{Chevillard2017} as we shall illustrate below. The goal of regularization and stationarization is to solve the two drawbacks of scale-invariant processes, and hence offer a ``more physical'' model for processes such as, e.g., fluid turbulence, to be compared with experimental data.

Fluid turbulence is an archetypal physical system that offers a perfect illustration. From the Kolmogorov 1941 perspective~\cite{Kolmogorov1991,Frisch:1995}, the Eulerian velocity field in homogeneous and isotropic turbulence presents a well-known scale-invariance property---the power spectral density evolves as a power law of the wavenumber with an exponent -5/3---within a 
%please confirm if change hyphen as minus sign, and please check all similar places
restricted region called the {\em inertial} range. In any experimental realization, for a finite Reynolds number, the inertial range corresponds to an interval of scales bounded from below by the {\em dissipative} scale and from above by the {\em integral} scale. Within the inertial range, the scale-invariance of turbulent velocity is well described by a Hurst exponent ${\cal H}=1/3$.

{Several approaches have been proposed to synthesize a stochastic process that has the same properties as the turbulent velocity, as can be seen for example in~\cite{Dimitriadis:2018} and the references therein. Of particular interest for us is the explicit introduction of both a dissipative and an integral scale in order to have a bounded inertial range, which can be performed by implementing the convolution of a white noise in several ways. We choose in the following to analyze two specific stochastic processes where a dynamical stochastic equation and explicit analytical comparison with fluid turbulence are available: the first one is a regularized and stationarized fBm and the second one is a regularized fractional Ornstein--Uhlenbeck process~\cite{Chevillard2017}.}
For consistency, we fix all along this section the small-scale $\epsilon=4$ and the large-scale ${\cal T}=\exp(9)=8103$.
For each process under consideration, we first generate a very long realization with $2^{23}$ data points and then divide it into segments of size $T=2^{16}$ points over which we estimate our quantities using scales $\tau$ in a logarithmic range between 1 and $2^{10}$. In order to analyze larger scales, we also down-sample the initial realization by a factor of 4, 16 and 64, and again perform the estimation on segments of the same size $T=2^{16}$ points.

%  (the same size as the experimental modane data) . . 

%%%%%%%%%%%%%%%%%%%%%%%%%%%%%%%%%%%%%%%%%%%%%%%%%%%
\subsection{Regularized and Stationarized fBm}

\begin{figure}[htb]
\includegraphics[width=.9\linewidth]{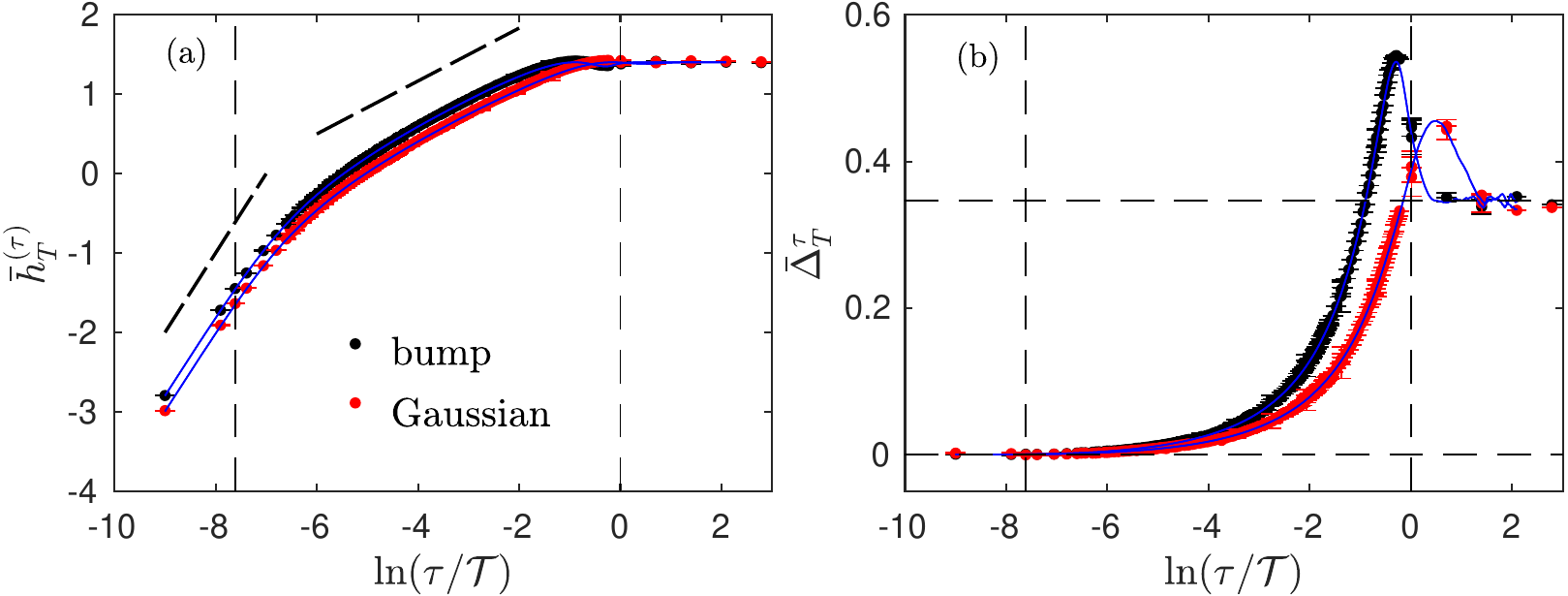}
\caption{{\bf Regularized and stationarized fBm.} 
Entropy rate $\bar{h}_T^{(\tau)}$ (\textbf{a}) and index $\bar{\Delta}_T^{\tau}$ (\textbf{b}) for two regularized and stationarized fBm with the same Hurst exponent $\mathcal{H}=1/3$ but with two different large-scale windows: bump (black {dots with error bars}) and Gaussian (red {dots with error bars}).
The blue continuous curves {(without error bars)} are approximations obtained by using only the correlation function (numerically estimated on the realizations) and Formulas (\ref{eq:G:h}) and (\ref{eq:G:Delta}).
The dashed vertical lines correspond to the dissipative scale $\epsilon=4$ and the integral scale ${\cal T}=\exp(9)=8103$ used to synthesize the process. 
The dotted straight lines in (\textbf{a}) are guides for the eyes with slopes $1$ and $1/3$. 
The horizontal dashed lines in (\textbf{b}) correspond to the special values $0$ and $\ln\sqrt{2}$.
}
\label{fig:fbmRegulBump1}
\end{figure}

We present in this section the results obtained with the regularized and stationarized fBm $B_{\epsilon, {\cal T}}$, a stochastic process introduced in \cite{Pereira2016}. This process has Gaussian statistics and perfectly mimics an fBm---with a prescribed exponent ${\cal H}$~---in a finite range of time-scales. However, contrary to the fBm, it has a finite second-order structure function at the large scale, larger than ${\cal T}$ while its power spectrum behaves as a power law with exponent -3---corresponding to a Hurst exponent 1~---at scales smaller than $\epsilon$. 
This process is generated as the convolution of a Gaussian white noise with the product:  
$
%B_{\epsilon,l} = bb * \left(
\left( \frac{t}{\sqrt{t^2+\epsilon^2}}\right)^{3/2-{\cal H}}.W_{\cal T}(t) 
%\right) 
\,,
$
where $W_{\cal T}$ is a large-scale function that insures stationarization~\cite{Pereira2016}. Among possible functions $W_{\cal T}$, we have used both the ``bump'' function $W_{\cal T}(t)=\frac{2\cal T}{a\sqrt{\pi}}\exp(-t^2/({\cal T}^2-t^2)$ for $|t|<{\cal T}$, $=0$ elsewhere, with $a=U(1/2,0,1)\simeq 0.603$ a particular value of the confluent hypergeometric function that ensures the normalization of $W_{\cal T}$,
and the Gaussian function $W_{\cal T}(t)=1/\sqrt{2\pi {\cal T}^2}\exp(-t^2/(2{\cal T}^2))$. 
Figure~\ref{fig:fbmRegulBump1} shows our findings for the two corresponding processes with $\mathcal{H}=1/3$.

The entropy rate $\bar{h}_T^{(\tau)}$ evolution with the time scale $\tau$ (Figure~\ref{fig:fbmRegulBump1}a) reveals three different regimes, as would the power spectrum~\cite{GBelinchon2016}.
Between the small and the large scales, indicated by vertical dashed lines, the entropy rate evolves linearly in $\ln\tau$ with a slope $\mathcal{H}=1/3$, just as it would have for a traditional fBm: this is the inertial regime.
For smaller scales below the dissipative scale, the entropy rate evolves faster, signaling the effect of the regularization: the slope is approximately +1 and the process is increasingly organized as the scale $\tau$ is reduced.
For larger scales, above the integral scale ${\cal T}$, the entropy rate is maximal and does not evolve with $\tau$: the process is then the most disorganized.
The transition from one regime to another is not sharp and it is difficult to recover the dissipative and integral scales by looking at the curve: both the effects of the regularization and of the stationarization invade the inertial region.

The index $\bar{\Delta}_T^{\tau}$ offers a deeper insight into the evolution of the dynamics of the process across the scales.
For smaller scales, $\bar{\Delta}_T^{\tau}=0$, as if the process was highly non-stationary as  a time-integrated fBm would be.
For larger scales above ${\cal T}$, $\bar{\Delta}_T^{\tau}\simeq \ln\sqrt{2}$, the value obtained for uncorrelated stationary processes such as a Gaussian random noise, {i.e.}, an fGn with ${\cal H}=1/2$.
In the inertial range, the index evolves non-monotonically between these two regimes, with a noticeable excursion above $\ln\sqrt{2}$ as if there are negative correlations at scales about the integral scale ${\cal T}$, before correlations vanishes at scales larger than the integral scale.

The evolution of the index $\bar{\Delta}_T^{\tau}$ thus suggests that the process evolves from a highly non-stationary process at a smaller scale to a stationary process at larger scales. Again, the transition between regimes is not sharp, but the effects of regularization and the stationarization are clearly visible, especially in comparison to the set of results for the fGn, fBm and time-integrated fBm presented in Figure~\ref{fig:FBM}.

%%%%%%%%%%%%%%%%%%%%%%%%%%%%%%%%%%%%%%%%%%%%%%%%%%%
\subsection{Regularized Fractional Ornstein--Uhlenbeck Process}

In this section, we present the results obtained with a regularized fractional Ornstein--Uhlenbeck process~\cite{Chevillard2017}.
This Gaussian process is an extension of the Ornstein Ulhenbeck process which exhibits scale invariance with a Hurst exponent ${\cal H}$ in a range of time scales. The relaxation coefficient $1/{\cal T}$ in its stochastic equation defines the integral scale ${\cal T}$ while an ad hoc regularization is introduced at small scale $\epsilon$~\cite{Chevillard2017}. For scales smaller than $\epsilon$,
the power spectrum of the process behaves as a power law with exponent -2, corresponding to a Hurst exponent 1/2. 

\begin{figure}[htb]
\includegraphics[width=.9\linewidth]{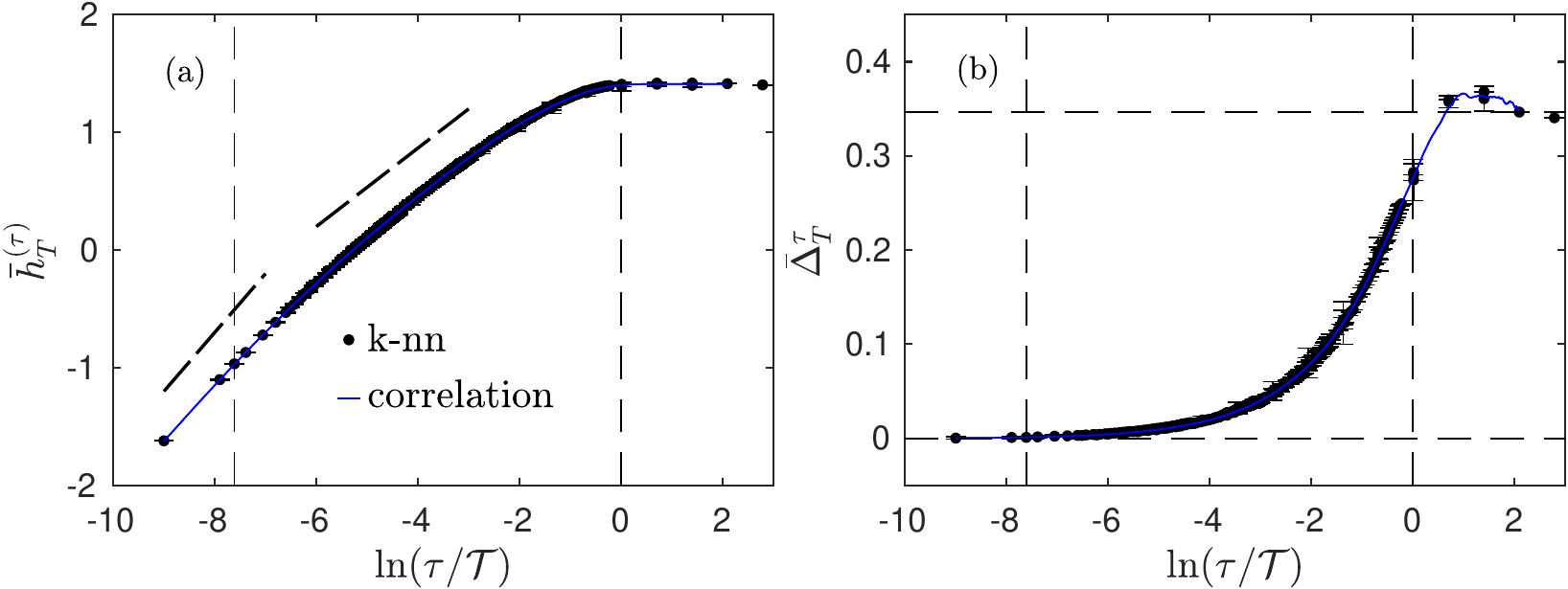}% This is a *.eps file
\caption{{\bf Regularized fractional Ornstein--Uhlenbeck.} 
Entropy rate $\bar{h}_T^{(\tau)}$ (\textbf{a}) and index $\bar{\Delta}_T^{\tau}$ (\textbf{b}) for a regularized Ornstein--Uhlenbeck process with $\mathcal{H}=1/3$. 
The black dots and error bars are obtained by directly  computing the information theory quantities on realizations of the process, while the blue continuous curves are obtained by using analytical Formulas (\ref{eq:G:h}) and (\ref{eq:G:Delta}) using only the correlation function, which was numerically estimated for the same realizations of the process.
The dashed vertical lines correspond to the dissipative scale $\epsilon=4$ and the integral scale ${\cal T}=8103$ used in the construction of the process.
The dotted straight lines in (\textbf{a}) are guides for the eyes with slopes $1/2$ and $1/3$. 
The horizontal dashed lines in (\textbf{b}) correspond to the special values $0$ and $\ln\sqrt{2}$.
}
\label{fig:Oulenbeck1}
\end{figure}

Figure~\ref{fig:Oulenbeck1} reports our findings for such a process with $\mathcal{H}=1/3$. Because the process is Gaussian, and its increments are Gaussian at all scales $\tau$, we can also estimate its entropy rate $\bar{h}_T^{(\tau)}$ and its index $\bar{\Delta}_T^{\tau}$ using Equations (\ref{eq:G:h}) and (\ref{eq:G:Delta}) in which we insert a numerical estimation of its correlation function; the corresponding estimations are reported in blue in Figure~\ref{fig:Oulenbeck1}.
We note that both the entropy rate and the index are very well estimated using the correlation function only when compared to the full estimation involving combinations of entropies.

The evolution of the entropy rate $\bar{h}_T^{(\tau)}$ with $\ln \tau$ (Figure~\ref{fig:Oulenbeck1}a) is very similar to the one observed for the regularized and stationarized fBm (Figure~\ref{fig:fbmRegulBump1}a), albeit the slope in the small scales region is different: it is close to +1/2, as expected, instead of +1 as for the fBm. 
The slope in the inertial range is again given by ${\cal H}=1/3$, and a constant value is reached for scales larger than the integral scale, albeit a little lower than the one for the stationarized fBm.

The index $\bar{\Delta}_T^{\tau}$ presents a behavior similar to that of the stationarized fBm: it increases from 0 to $\ln\sqrt{2}$, but the increase seems monotonic for the Ornstein--Uhlenbeck, or with a much smaller overshoot before reaching the constant value $\ln\sqrt{2}$. 

%%%%%%%%%%%%%%%%%%%%%%%%%%%%%%%%%%%%%%%%%%%%%%%%%%%
\section{Fully Developed Fluid Turbulence}
\label{sec:turbulence}

In this section, we analyze the experimental fluid turbulence in various experimental setups. As evoked in Section~\ref{sec:physProc}, fluid turbulence is the physical archetypal system where a power law spectrum is observed in an {\em inertial} range, in between a {\em dissipative} scale and an {\em integral} scale.
While the fBm (Section~\ref{sec:fGnfBm}) with the Hurst exponent 1/3 is a classical model for the inertial range only~\cite{Kolmogorov1991,Frisch:1995}, regularized and stationarized fBm as well as regularized fractional Ornstein--Uhlenbeck process (Section~\ref{sec:physProc}), both offer more realistic models by including the {\em dissipative} and {\em integral} scales in addition to the inertial range. We now want to compare these two models with experiments, especially with regard to our new index.

We use two sets of Eulerian longitudinal velocity measurements which have been previously characterized in detail.
The first dataset was obtained in a grid setup, in the Modane wind tunnel~\cite{Kahalerras1998}. The sampling frequency of the setup was 25 kHz, the mean velocity of the flow is $\left\langle v\right\rangle=20.5$ m/s, and the Taylor-scale based Reynolds number of the flow is approximately $R_\lambda=2700$, large enough for the flow to be considered as exhibiting fully developed turbulence. 
For this dataset, we use the Taylor frozen turbulence hypothesis~\cite{Frisch:1995} in order to interpret temporal variations as spatial ones and we can then use the Bachelor model to estimate the larg- scale $L=0.74$ m corresponding to a large temporal scale ${\cal T}\equiv L/\left\langle v\right\rangle$ = 36 ms.
%; so we consider the temporal measurements as representing the spatial variations of the longitudinal flow velocity. Thus, while we examine in the following the temporal scales $\tau$, these are directly related to the spatial scales $l$ by $\tau=l/\langle v\rangle$. 
%
The second dataset was obtained from a helium jet setup~\cite{Chanal2000}. It consists of several experiments for various Taylor-scale based Reynolds numbers $R_\lambda=89$, $208$, $463$, $703$ and $929$. 
%Again, the Taylor hypothesis is used but with the instantaneous frequency instead of the mean velocity, which is equivalent to Lumley's prescription~\cite{J. Lumley, Phys. Fluids 8, 1056 (1965)}. In the following, we present our results using time-scales, although corresponding experimental data as well as turbulence theoretical approaches both implies the use of space-scales. This allows for a simpler comparison with results from former sections, although it should be noted that in this representation, every experiment appears as recorded at a different sampling rate.
%
For each experiment, we computed the integral scale $\cal{T}$ as the scale for which the index reaches the value corresponding to an uncorrelated Gaussian process, {i.e.}, $\bar{\Delta}_T^{(\tau={\cal T})}=\ln\sqrt{2}$.
We checked that this integral time scale ${\cal T}$ is in perfect agreement with the spatial integral time scale $L$ obtained from a fit of the Bachelor model, within the usual error bars, as reported in~\cite{Chanal2000}.

\begin{figure}[htb]
\includegraphics[width=.9\linewidth]{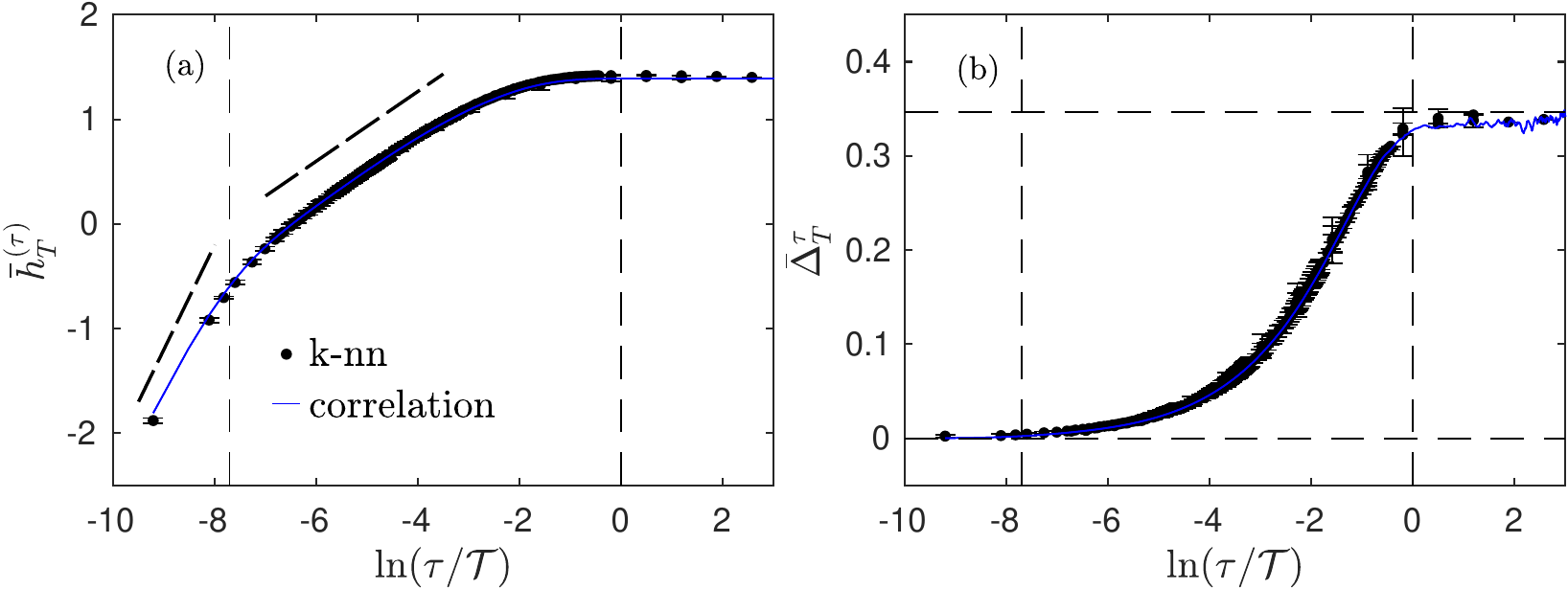}
\caption{{\bf Experimental grid turbulence at $R_\lambda$ = 2500.} 
Entropy rate $\bar{h}_T^{(\tau)}$ (\textbf{a}) and index $\bar{\Delta}_T^{\tau}$ (\textbf{b}) for Modane experimental velocity measures. 
Black dots with error bars correspond to the complete information theory measure with Equation (\ref{eq:Delta:MI}), while blue lines correspond to the estimates (\ref{eq:G:Delta}) only involving the correlation function.
The vertical dashed lines correspond to the dissipative $\epsilon=5$ and integral ${\cal T}=2530$ scales obtained by a fit of the Bachelor model~\cite{Bachelor1951}. 
The dotted straight lines in (\textbf{a}) are guides for the eyes with slopes $1$ and $1/3$. 
The horizontal dashed lines in (\textbf{b}) correspond to the special values $0$ and $\ln\sqrt{2}$.
}
\label{fig:Modane}
\end{figure}

\begin{figure}[htb]
\includegraphics[width=.9\linewidth]{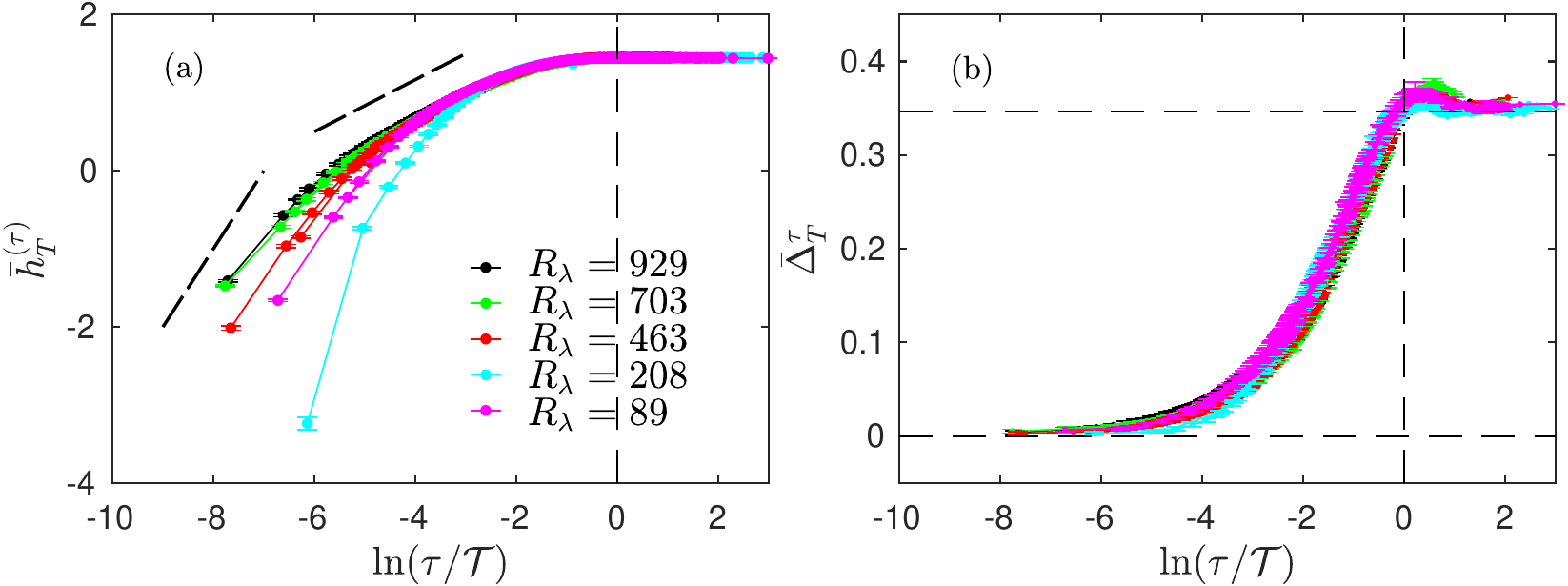}
\caption{{\bf Experimental jet turbulence at various $R_\lambda$.} 
Entropy rate $\bar{h}_T^{(\tau)}$ (\textbf{a}) and index $\bar{\Delta}_T^{\tau}$ (\textbf{b}), for the experimental velocity measures of (helium) jet turbulence at Reynolds 929 (black), 703 (blue), 463 (red), 208 (magenta) and 89 (cyan). The velocity signals are normalized ($\sigma=1$). 
The dotted straight lines in (\textbf{a}) are guides for the eyes with slopes $1$ and $1/3$. 
The horizontal dashed lines in (\textbf{b}) correspond to the special values $0$ and $\ln\sqrt{2}$.
}
\label{fig:jet}
\end{figure}

To characterize the velocity datasets, we computed their entropy rate $\bar{h}_T^{(\tau)}$ as well as their index $\bar{\Delta}_T^{\tau}$, as the functions of the scale expressed with the non-dimensional ratio $\tau/{\cal T}$. The results are presented in Figure~\ref{fig:Modane} for the Modane experiment and in Figure~\ref{fig:jet} for the helium jet experiments.

We first examined the Modane experiment which has a large Reynolds number.
In\linebreak Figure~\ref{fig:Modane}a, we clearly see that the entropy rate reveals the three domains of scales described by the Kolmogorov theory~\cite{Kolmogorov1991}. $\bar{h}_T^{(\tau)}$ behaves as a power law with an exponent close to $1$ in the dissipative domain, and with an exponent close to $1/3$ in the inertial domain, while it reaches a plateau when entering the integral domain. Vertical dashed lines in Figure~\ref{fig:Modane}a indicate the dissipative and integral scales as obtained with the Bachelor model~\cite{Bachelor1951}. In\linebreak Figure~\ref{fig:Modane}b, we see that the index $\bar{\Delta}_T^{\tau}$ evolves smoothly and monotonically from $0$ at small scales, up to $\ln\sqrt{2}$---the value for a stationary an uncorrelated Gaussian process---at large scales. 

It thus seem that, although the behavior of the entropy rate of the experimental fluid turbulence (Figure~\ref{fig:Modane}a) is better described by the regularized and stationarized fBm model (Figure~\ref{fig:fbmRegulBump1}a), the behavior of the index (Figure~\ref{fig:Modane}b) bears greater resemblance to that of the fractional Ornstein--Uhlenbeck process (Figure~\ref{fig:Oulenbeck1}b).

We then examined the influence of the Reynolds number by studying the helium jet experiments.
In Figure~\ref{fig:Modane}, we see that both the entropy rate $\bar{h}_T^{(\tau)}$ (Figure~\ref{fig:jet}a) and the index $\bar{\Delta}_T^{\tau}$ (Figure~\ref{fig:jet}b) both behave as in the Modane experiment. 

Let us first describe the evolution of the entropy rate with $\ln(\tau/\cal{T})$ from the large scales down to the smaller scales. 
For all Reynolds numbers, $\bar{h}_T^{(\tau)}$ is maximal and constant in the integral domain, while it  linearly decreases with a slope $1/3$ in the inertial range. For smaller scales below the dissipation scale, the entropy rate linearly  decreases with a slope of approximately $1$. As expected, when the Reynolds number is increased, the dissipation scale is smaller, and the inertial range is thus wider~\cite{Frisch:1995}.

We now describe the evolution of the index $\bar{\Delta}_T^{\tau}$ with $\ln(\tau/\cal{T})$. Again, the index varies from 0 at small scales to $\ln\sqrt{2}$ at large scales, but all curves for all Reynolds numbers now seem to overlap. In particular, the dissipative scale does not seem to play a particular role in the behavior of the index. This may suggest that this quantity only probes the transition from the inertial range to the integral domain, {i.e.}, the changes in the stationarity at the scale $\tau$. Interestingly, we see that the index $\bar{\Delta}_T^{\tau}$ slightly overshoots the value $\ln\sqrt{2}$ around the integral scale, before converging to this value from above for larger values of $\tau$. This behavior is more pregnant in experiments at $R_\lambda=208$ (magenta) and $R_\lambda=703$ (dark blue), and less obvious in the other ones. The transition of the index from $0$ to $\ln\sqrt{2}$ may not be monotonic, and thus similar to what was observed for the regularized and stationarized fBm (Figure~\ref{fig:fbmRegulBump1}b) and the regularized fractional Ornstein--Uhlenbeck process (Figure~\ref{fig:Oulenbeck1}b); but in that respect, the behavior of the experimental jet data bear greater resemblance to  that of the regularized fractional Ornstein--Uhlenbeck process.

\begin{figure}[htb]
\includegraphics[width=.9\linewidth]{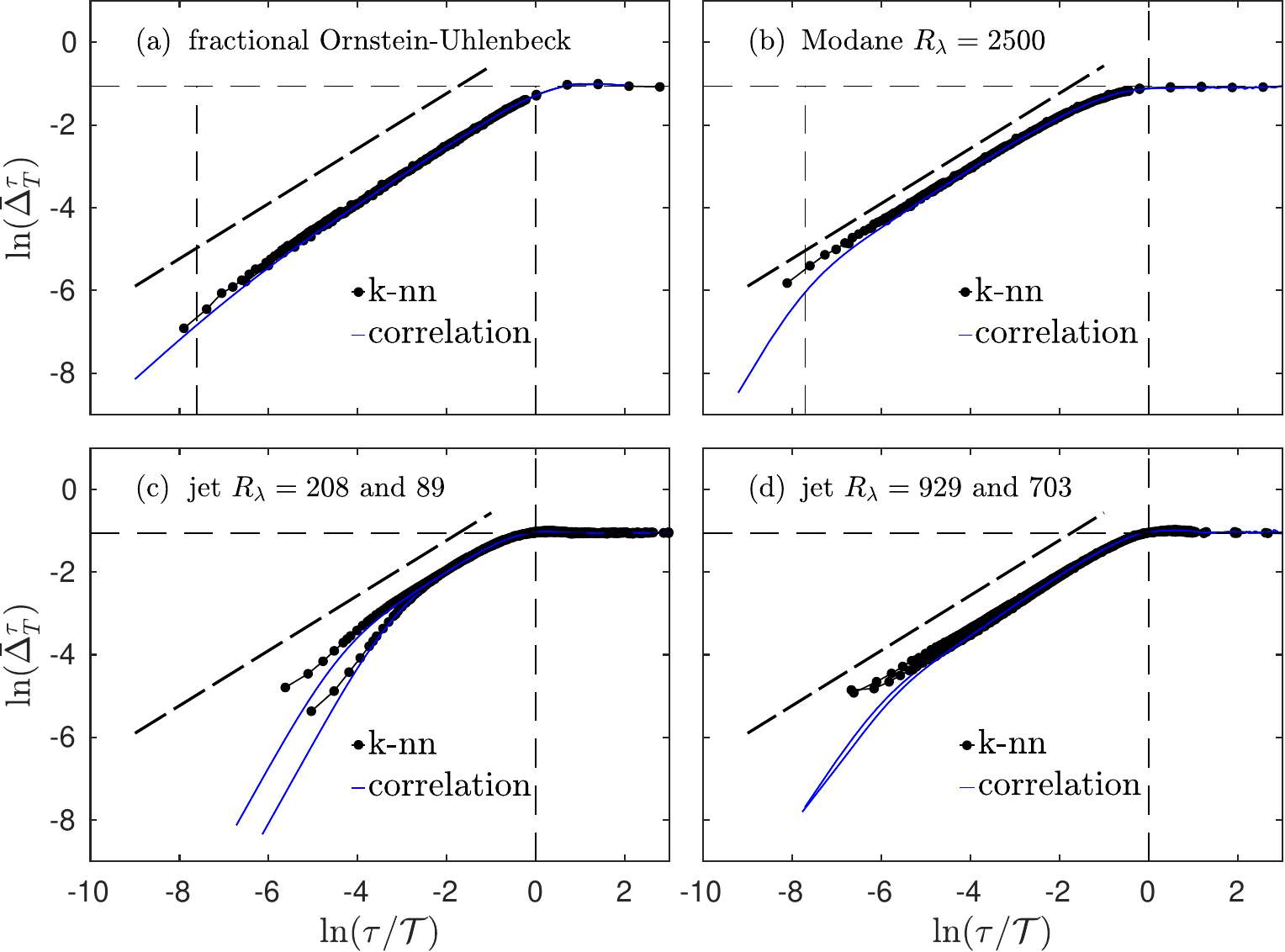}
\caption{{\bf Dependences of several processes.} 
Logarithm $\ln\left( \bar{\Delta}_T^{\tau} \right)$ of the index for the fractional Ornstein--Uhlenbeck process (\textbf{a}), grid turbulence in the Modane wind-tunnel (\textbf{b}), helium jet turbulence at smaller (\textbf{c}) and larger (\textbf{d}) Reynolds numbers.
{Thick black dots represent the complete information theory measure from Equation (\ref{eq:Delta:MI}), while blue lines correspond to the estimate (\ref{eq:G:Delta}) only involving the correlation function}. 
The horizontal dashed line corresponds to $\bar{\Delta}_T^{\tau}=\ln\sqrt{2}$.
The vertical dashed lines correspond to the dissipative ((\textbf{a},\textbf{b}) only) and integral scales.
The thick dashed line is a guide for the eye with a slope $2{\cal H}=2/3$.}
\label{fig:All}
\end{figure}

In order to better apprehend what occurs around the integral scale and around the dissipative scale, we plot in Figure~\ref{fig:All} the logarithm of the index $\bar{\Delta}_T^{\tau}$, as a function of $\ln(\tau/\cal{T})$, for the fractional  Ornstein--Uhlenbeck process and experimental longitudinal velocity measurements.

Together with the estimation using the information theoretical definition (\ref{eq:Delta:MI}) (black dots), we also plot the simpler estimation that  only uses the correlation function and formula (\ref{eq:G:Delta}) (blue line). This last measurement is only supposed to match the real estimation when the process is Gaussian and stationary, which is the case for the fractional  Ornstein--Uhlenbeck: as can be seen in Figure~\ref{fig:All}a), both estimates are indeed very close for all time scales.
For experimental data, the agreement is very good at larger scales, from the inertial domain up to the integral domain, but a very noticeable deviation appears at smaller scales.

Let us first focus on the Modane experiment, which has the largest Reynolds number, to describe what happens at smaller scales.
As observed in Figure~\ref{fig:Modane}a), the entropy rate is very well approximated for all scales by Equation (\ref{eq:G:h}) which uses the correlation only. For the index, the discrepancies at smaller scales may thus be expected to arise from the entropy of the increments according to Equation (\ref{eq:Delta:diff}). It is important to remember that the statistics of the increments are Gaussian at larger scales only, about the integral scale and larger, while they are more and more non-Gaussian at smaller scales; this phenomena is referred to as the {\em intermittency} of turbulence. 
The deviation from Gaussian statistics has previously been studied~\cite{GraneroBelinchon2018} by measuring the extra information in the entropy of the increments, with respect to the entropy that can be estimated by assuming purely Gaussian statistics and using the standard deviation only. The presence of intermittency therefore leads to a larger value of the index compared to what can be estimated using only the correlation function. The difference between the two estimates should correspond to the Kullback--Leibler divergence introduced in~\cite{GraneroBelinchon2018}. We note that only the index---in its complete information theoretical form~---probes higher-order statistics and the full dependences of the process, whereas the correlation estimate (\ref{eq:G:Delta}) only takes into account the second-order moment and correlations.

Looking at the behavior of the index for smaller scales, we also observe that there is no clear influence of the dissipative scale. 
Even after taking the logarithm---so even when enlarging the perspective on the smallest values of the index---the index seems to behave exactly the same in the inertial range and in the dissipative range, as a power law of the scale.
The exponent of the power law can be derived, using the approximation (\ref{eq:G:Delta:exp}) for small correlation and assuming a Gaussian process with a power-law scaling of the variance of the increments; we then expect the exponent of the power law to be $\zeta_2=2{\cal H}$ for a scale-invariant process. The thick dashed black line in all panels of Figure~\ref{fig:All} represents this exponent $2{\cal H}=2/3$ and shows that it offers a good approximation for all the processes under study here.

It is worth recalling that turbulence data are usually considered stationary, but this consideration is made at larger scales. A very local observation, {i.e.}, considering smaller scales or examining a short portion of the velocity field, usually reveals a non-stationary process, in the form of local trends that eventually compensate when averaged over many short portions, hence over longer scales. This scale-dependent non-stationarity is measured by the index, and we interpret the difference between the index and its Gaussian approximation as an increase in non-stationarity due to the full dependence structure of the process.

%%%%%%%%%%%%%%%%%%%%%%%%%%%%%%%%%%%%%%%%%%%%%%%%%%%
\section{Discussion and Conclusions}
\label{sec:conclusion}

Using information theory, we proposed an index $\bar{\Delta}_T^{\tau}(X)$ which is a good candidate to quantify the non-stationarity of a process at a given scale $\tau$.
This index is defined for a discrete-time process $\{x_t\}_{t\in {\mathbb N}}$ as the difference between the information contained in the increment $\delta_{\tau}x_t=x_t-x_{t-\tau}$ at scale $\tau$ and the new information in $x_t$ that was not already present in $x_{t-\tau}$.
By varying the scale $\tau$, the index allows a multi-scale characterization of the process.

The index takes real positive values. {For Gaussian processes,} a value of $\ln\sqrt{2}$ indicates stationarity, and lower values indicate some non-stationarity. The index saturates at zero for non-stationary processes, so the non-stationarity degree cannot be measured directly. Nevertheless, we showed using the fGn and its successive time-integrations that iteratively time-deriving  the signal (or iteratively taking time-increments) and counting the number of iterations required to obtain values of the index close to $\ln\sqrt{2}$ should be enough to infer the integer part of the non-stationary degree.
{This methodology holds for non-Gaussian processes, although the very value $\ln\sqrt{2}$ for the constant might depend on the shape of the large-scale probability density function; we are currently investigating such processes which are not Gaussian at larger scales, and correspond to non-physical processes within our approach.}

We showed that, for physically sound processes which are stationary at larger scales, the index is not only able to reveal at which scales larger or about the integral scale ${\cal T}$ the process is indeed stationary, but also to quantify how the process becomes non-stationary when the scale $\tau$ is reduced. Using synthetic data as well as experimental velocity recordings in fluid turbulence, we showed that the index contains information that is not grasp by the correlation function alone, and because of its very definition, the index probes the full dependence structure of the process. We thus note that for a process to qualify as stationary, its index at larger scales (corresponding to the size of the observation time-window) must approach the value $\ln\sqrt{2}$, which implies that not only the correlations but also all dependences are vanishing {while the distribution becomes more and more Gaussian when the scale is increased}.
It is worth noting that using the criterion $\bar{\Delta}_T^{\tau={\cal T}_\Delta}=\ln\sqrt{2}$ to define the (large) scale ${\cal T}_\Delta$ at which all dependences have vanished leads to an integral scale estimation that is always larger than the integral scale ${\cal T}$ imposed in synthetic processes (Figures~\ref{fig:fbmRegulBump1} and \ref{fig:Oulenbeck1}), or larger than the integral scale ${\cal T}$ obtained from a fit of the Bachelor model (Figure~\ref{fig:Modane}b). This is not surprising as the integral scale ${\cal T}$ indicates the typical location of the boundary between the inertial and integral domains, and so it corresponds to a region where both inertial and integral behaviors are overlapping, and some remaining dependences from the inertial range are expected to exist.

Additionally, the index does not distinguish between the inertial and dissipative domains, whereas the correlation and the power spectrum density both do. 
For scale-invariant processes with stationary increments and noting ${\cal H}$ the Hurst exponent, the behavior of the index $\bar{\Delta}_T^{\tau}(X)$ with the scale $\tau$ is very close to a power law with the exponent $2{\cal H}$. We suggested that this property generalizes to multifractal processes where we expect the index to behave as a power law with the exponent $\zeta_2$.

As illustrated with scale-invariant processes, the non-stationarity is directly related to the roughness measured by the Hurst exponent ${\cal H}$. The ersatz entropy rate $\bar{h}^{(\tau)}_T$ also offers a way to assess the Hurst exponent---which can be estimated as the slope of the linear evolution of $\bar{h}^{(\tau)}_T$ with $\ln\tau$---but this requires a process with stationary increments~\cite{GraneroBelinchon2019}, so $0<{\cal H}<1$, as can be seen in the second line of Figure~\ref{fig:FBM} where it only works for the fBm. For processes with ${\cal H}\ge 1$, the slope of the linear evolution of $\bar{h}^{(\tau)}_T$ with $\ln\tau$ saturates at the value 1, and successive time-derivation are then required to measure the (non-integer part of the) Hurst exponent. On the contrary, the index can be estimated on any process, and the comparison with the special value $\ln\sqrt{2}$ always holds, albeit eventually following the iterative recipe above. Because the presence of a dissipative range changes the slope of $\bar{h}^{(\tau)}_T$ with $\ln\tau$, whereas it does not appear to change the slope of $\ln \bar{\Delta}_T^{\tau}(X)$, it suggests that the index is a better tool to probe the non-stationarity.

The index is closely related to both the ersatz entropy rate~\cite{GraneroBelinchon2019} and the Kullback--Leibler divergence~\cite{GraneroBelinchon2018}. Just like these two quantities, the index offers a novel perspective on fluid turbulence or on any stochastic process by providing a new insight on its regularity and stationarity properties, as a function of the scale.
Future work is required to fully understand how these three information theoretical quantities quantitatively relate in the time-averaged framework for non-stationary processes.

%%%%%%%%%%%%%%%%%%%%%%%%%%%%%%%%%%%%%%%%%%
%\authorcontributions{Conceptualization, N.B.G.; Investigation, C.G.B., S.G.R. and N.B.G.; Methodology, S.G.R. and N.B.G.; Visualization, C.G.B. and S.G.R.}

%\funding{This research received no external funding.}

%\conflictsofinterest{The authors declare no conflict of interest.}

%\begin{acknowledgments}
%This work was supported by the LABEX iMUST (ANR-10-LABX-0064) of Universit\'e de Lyon, within the program "Investissements d'Avenir" (ANR-11-IDEX-0007) operated by the French National Research Agency (ANR).
%\end{acknowledgments}

%\bibliographystyle{unsrt}
%\bibliography{THEBIBLIO}

%\tableofcontents

\end{document}